\documentclass[11pt]{article}
\usepackage{moriond,epsfig}

\def\be{\begin{equation}}
\def\ee{\end{equation}}
\def\bea{\begin{eqnarray}}
\def\eea{\end{eqnarray}}

\def\wg{$W\gamma$ }

\def\pythia{\textsc{Pythia}} 
\def\madgraph{\textsc{Madgraph}} 
\def\ptg{$p_{T}^{\gamma}$ } 
\def\pbi{~pb$^{-1}$ } 
\def \et{E_{T} }
\def\met{$\not\!\!\et$ }
\def \sihih{$\sigma_{\eta \eta}$ } 

\def \etg{E_{T}^{\gamma} }
\def \ptg{$p_{T}^{\gamma}$ }

\graphicspath{{plots/}}

\begin{document}

\title{Study of \wg events at the CMS with 7~TeV LHC data} 

\author{D. Majumder, on behalf of the CMS collaboration.}  
\address{Tata Institute of Fundamental Research, \\ 
1 Homi Bhabha Road, Mumbai 400 005, India. 
} 

\maketitle

\abstract{The measurement of the inclusive cross section for \wg production is presented based on 36\pbi of data acquired with the CMS detector from 7~TeV LHC collisions in 2010. Comparisons are made with the predictions of the standard model. The $W$ bosons are identified through their leptonic decays to electrons and muons. The \wg cross section is sensitive to anomalous triple-gauge couplings and hence this measurement probes physics beyond the standard model.} 

\section{Introduction}

Diboson productions ($W\gamma$, $WW$, $WZ$, $Z\gamma$, $ZZ$)  are at the frontier of the standard model (SM) physics to be studied at the LHC  before embarking on the search for physics beyond the standard model. In particular, the production rate of the \wg process is large enough to be measured with data from the first year of the running of the LHC at a centre-of-mass energy of 7~TeV. The presence of anomalous $WW\gamma$ coupling (ATGC) modifies the cross section and the photon transverse momentum (\ptg) distribution and hence the measurement of the cross section is the first step towards the determination of the $WW\gamma$ coupling values. Here we present the first measurement of the cross section of \wg production at $\sqrt{s} = 7$~TeV, using 36\pbi of data collected by the CMS detector in 2010. The measurement was done in the electron and muon decay modes of the $W$-boson. 

\section{\wg event selection}
\label{sec:evtSel} 

The CMS detector and its trigger and data acquisition system~\cite{:2008zzk} was used to record events from the LHC proton-proton collisions. The selection of potential \wg events was done using single electron or single muon triggers, which require the presence of at least one electron or a muon with a transverse momentum above a given threshold, in the event. The \wg events are studied in the final state $\ell\nu\gamma$, where $\ell$ represents either an electron or a muon. 

The \wg production cross section is measured within the phase space defined by $E_{T}^{\gamma}>10$~GeV and $\Delta R(\ell,\gamma)>0.7$ where $E_{T}^{\gamma}$ is the transverse energy of the photon in the \wg final state and $\Delta R(\ell,\gamma)\equiv\sqrt{(\eta_{\ell}-\eta_{\gamma})^{2} +(\phi_{\ell}-\phi_{\gamma})^{2}}$. $\eta_{\ell}$ and $\eta_{\gamma}$ are the pseudorapidities of the lepton and the photon respectively and $\phi_{\ell}$ and $\phi_{\gamma}$ are their azimuthal angles. 

The main background to the detection of \wg events are W+jets, where the final state contains a $W$-boson and the photon is faked by the jet. This background is estimated from data using the methods described in Section~\ref{sec:bkgEstimate}. The lesser backgrounds are due to $t\bar{t}$ events, multijet QCD processes and other diboson events. These backgrounds are determined using event samples from Monte Carlo event generators. The W+jets, Z+jets and $t\bar{t}$ events were generated using a combination of the \madgraph~\cite{Maltoni:2002qb} and \pythia~\cite{Sjostrand:2006za} event generators while the rest were generated using only \pythia. Monte Carlo generated samples were processed using a \textsc{GEANT4}-based~\cite{Agostinelli:2002hh} simulation of the CMS detector and reconstructed in the same way as those from the collision data. For the background determination, all Monte Carlo samples are normalized to the integrated luminosity of the collision dataset and next-to-leading order cross section predictions were used. 

The electrons are reconstructed in the electromagnetic calorimeter (ECAL)~\cite{EGM-10-004} and are identified using two sets of criteria, one based on the electron shower shape in the ECAL and the other based on the spatial matching of the charged track to the cluster of energy deposited in the ECAL. These selection criteria are designed to have a good rejection for jets from QCD multijet processes where the jet may be misidentified as an electron. The selection efficiency is about 80\% for electrons from the decay of the $W$-boson. Electrons originating from the pair production by photons interacting in the material of the detector are suppressed by the CMS electron reconstruction algorithm~\cite{EGM-10-005}. 

The muon reconstruction in CMS utilises information from both the muon chambers as well as the silicon tracker, to build up track segments which are finally matched to produce a muon candidate. For a well-reconstructed muon, its track should have at least 11 hits in the silicon tracker and should originate from the primary vertex in the event. The muon selection criteria has an efficiency of 95\%.

The electrons and the muons are further required to be isolated, with energy deposits in the surrounding detector elements that are below required values. Both the electron and the muon selection criteria used in this analysis follow the standard selection used for the measurement of the $W$ and $Z$ boson cross section in CMS~\cite{Khachatryan:2010xn}. The $W$-boson candidates are reconstructed using a well-reconstructed charged lepton and the missing transverse energy (\met), due to the neutrino from W-boson decay, in the event. Both the electrons and muons in the final states are required to have a transverse momentum above 20~GeV/c. The electrons' and the muons' pseudorapidities should be $|\eta_{e}| <2.5$ and $|\eta_{\mu}| < 2.1$, respectively. The \met in the detector is reconstructed using the particle-flow (PF) method~\cite{PFT-09-001}, which aims to reconstruct every particle in an event by combining the information from all CMS subdetector systems. The particles reconstructed are the electrons, the photons, the muons and the charged and neutral hadrons. The PF \met is then evaluated as the negative of the vector sum of the transverse momenta of all reconstructed particles in the event~\cite{JME-10-005}. For an  event to qualify as one containing a $W$-boson candidate, the value of PF \met should be above 25~GeV. 

The photon candidates are reconstructed as clusters of energy deposited in the ECAL with the photon pseudorapidity in the range $|\eta_{\gamma}| < 1.44$ or $1.57 < |\eta_{\gamma}| < 2.5$. The photon selection criteria is aimed at reducing fakes due to electrons, by imposing the requirement that there should not be any hits in the pixel detector pointing at the ECAL energy deposit. The ratio of photon energy deposit in the hadron calorimeter (HCAL), which lies just behind the ECAL, to that in the ECAL should be less than 0.05. The photon is also required to be isolated in the tracker and the calorimeters. Further, the electromagnetic shower profile in pseudorapidity must be consistent with that of a photon~\cite{EGM-10-005}. The photon's selection criteria are mostly geared towards strongly suppressing misidentified jets and has an efficiency of 90\%, while achieving a significant reduction in the number of the fake jets. 

\section{Determination of background}
\label{sec:bkgEstimate}

With the above selection criteria, 452 \wg events are selected in the electron channel, while in the muon channel 520 events are selected. The background contamination in these events are evaluated using both the data as well as the Monte Carlo simulations. Two complementary approaches are used to determine the fake photon background. 

\begin{figure} 
\begin{center}
\includegraphics[width=0.38\textwidth]{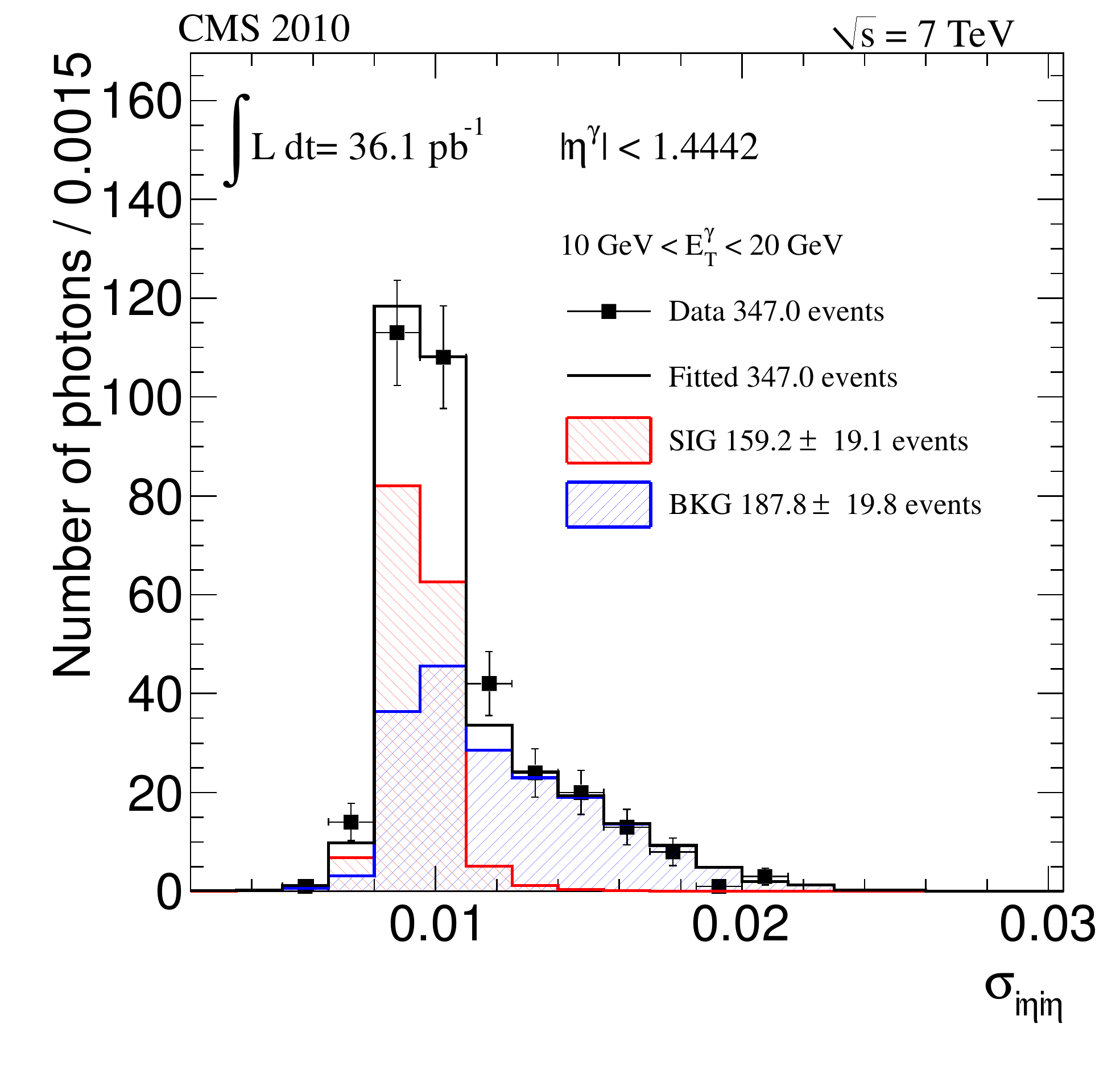}
\includegraphics[width=0.38\textwidth]{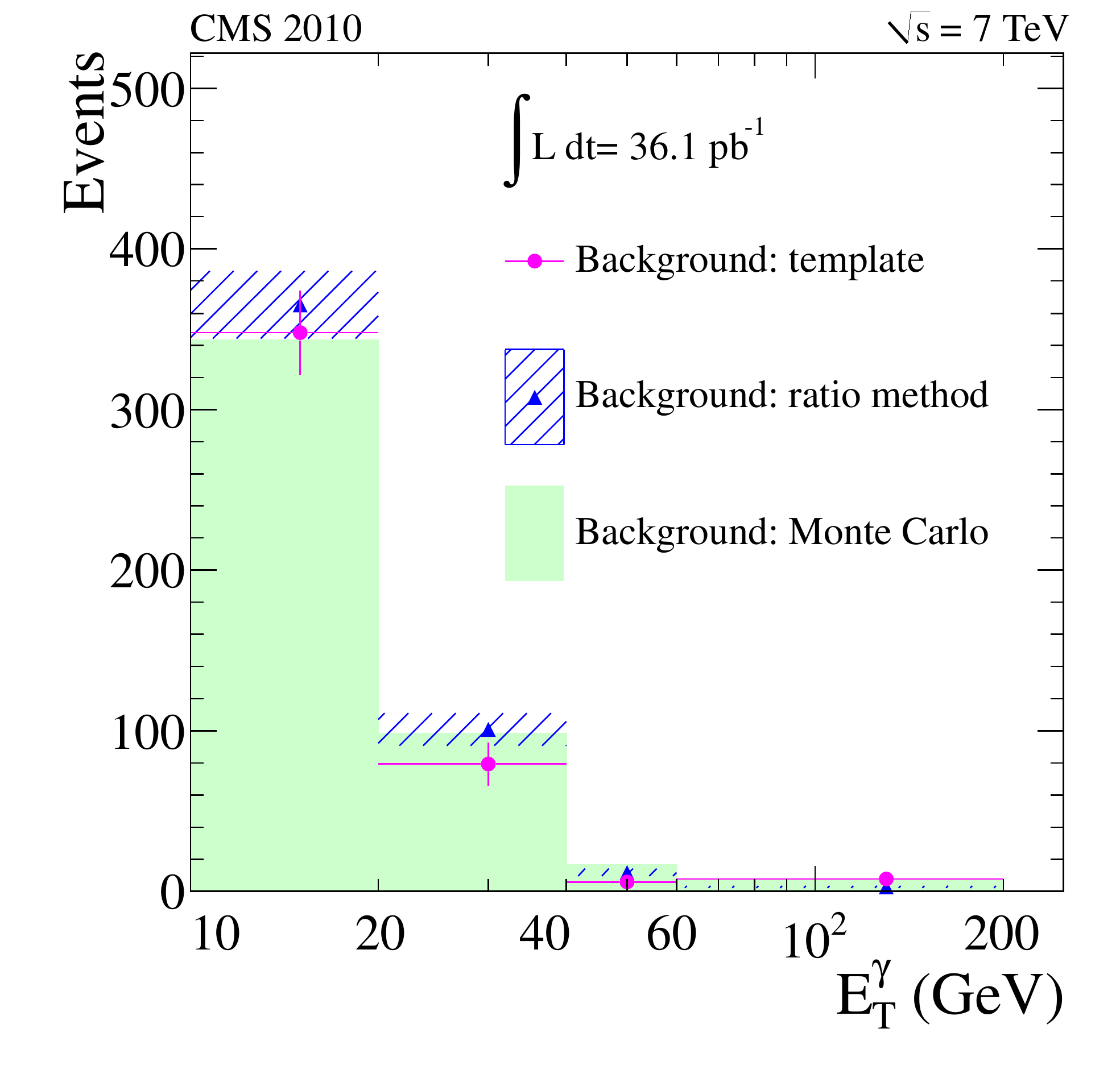}
\caption{The fit of the signal (true photon) and the background (fake photon) \sihih shapes to the data to extract the signal component (left) and the comparison of the background yields from the shape method, the ratio method and Monte Carlo prediction for the electron and the muon channels combined (right). Both the statistical and systematic uncertainties are included.}   
\label{fig:templFit}
\end{center}
\end{figure}

The first approach, the {\it template or shape method} uses the photon shower shape profile in pseudorapidity, denoted by \sihih, which describes the spread of the photon's electromagnetic shower in the pseudorapidity direction. Fake photons from jets have a different \sihih distribution from real photons and hence templates of \sihih for real and fake photons can be used to determine the background component in data. An extended maximum likelihood fit is used to obtain the signal component in the selected events, as shown in Fig.~\ref{fig:templFit} (left) for the muon channel, with a particular $\etg$ range of 10~GeV$<\etg<$20~GeV. The background yield using the shape method is 213.6$\pm$15.6~(stat.)~$\pm$23.9~(syst.) for the muon channel and 213$\pm$~16.1~(stat.)~$\pm$24~(syst.) for the electron channel. The systematic uncertainty is mainly due to the variation of the signal template shapes in data and Monte Carlo and the contribution of real photons in the background templates obtained from the QCD multijet events. 

The second method, the {\it ratio method}, is based on the assumption that the jets misidentified as photons in W+jets events have the same properties as the jets from QCD multijet events. The $\etg$-dependent ratio of the number of fake photons being isolated to that being non-isolated is determined from an independent QCD multijet sample in data, which is then folded in with the number of events with a $W$-boson and a non-isolated fake photon, which yields the number of W+jets events according to the relation 
\begin{center}
\begin{math}
N_{\rm W+jets} = \Big( \frac{N_{\rm isolated~\gamma}}{N_{\rm non-isolated~\gamma}} \Big)_{\rm QCD~multijet} \cdot N_{\rm W+non-isolated~\gamma} 
\end{math}
\end{center}
The non-isolated photon condition imposed is that the fake photon candidates should pass all the photon selection requirements listed in Section~\ref{sec:evtSel} but fails the tracker isolation requirement. The background estimated using the ratio method gives 260.5$\pm$18.7~(stat.)~$\pm$16.1~(syst.) events in the muon channel and 220.0$\pm$15.8~(stat.)~$\pm$13.9~(syst.) events in the electron channel. The systematic uncertainties involved are due to the modelling of the ratio distribution as well as due to the contamination of the real photons in the ratio obtained from the jet-triggered dataset. 

The agreement between the shape and the ratio methods are shown in Fig.~\ref{fig:templFit} (right). The two methods yield similar background estimates, which are also compared with the Monte Carlo prediction of the W+jets background. The ratio method has a smaller systematic uncertainty than the shape method and thus in the determination of the cross section, the background estimated using the ratio method is used. 

The smaller backgrounds that are measured directly from the Monte Carlo simulated datasets have systematic uncertainties on them mostly due to the electron, photon and muon energy scale uncertainties. 

\section{Estimation of the cross section}
\label{sec:cs}

\begin{figure} 
\begin{center}
\includegraphics[width=0.39\textwidth,height=0.3\textwidth]{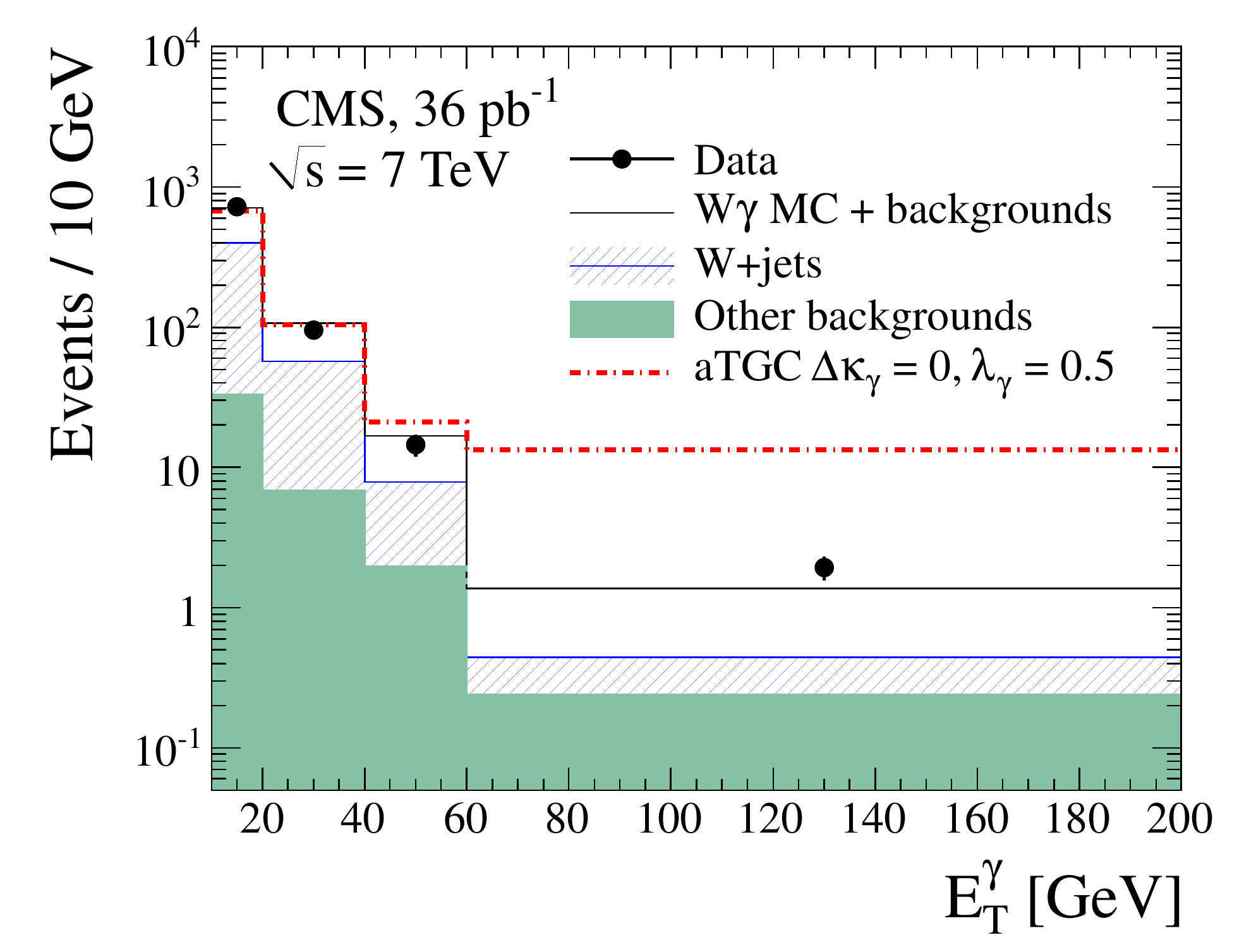}
\includegraphics[width=0.39\textwidth,height=0.3\textwidth]{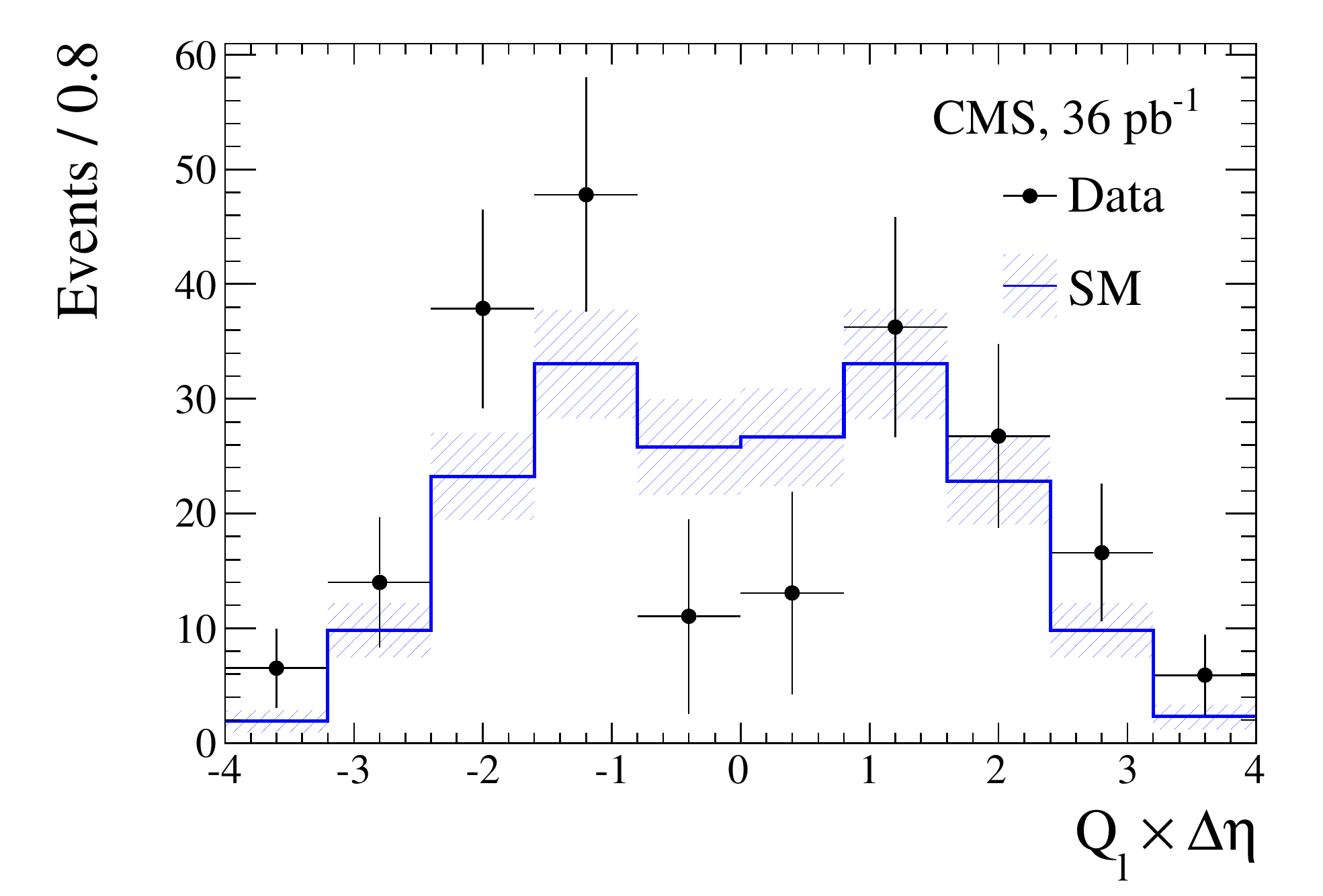}
\caption{Left: The transverse energy distribution for photons in the selected \wg events. The data are shown in black points with error bars while the expected SM \wg signal is shown as a black solid histogram. The blue hatched histogram are the fake photons from jets as measured using the ratio method. The other backgrounds like QCD, $t\bar{t}$ and dibosons are shown in green. The red dot-and-line histogram shows the $\etg$ distribution in the presence of a reference value of anomalous $WW\gamma$ coupling. Right: The background-subtracted charge-signed rapidity difference between the photon and the charged lepton, for the combined electron and muon channels of \wg production, is shown for data and SM simulation. The results of the Kolmogorov-Smirnov test of the agreement between data and MC prediction is 57\%, which indicates a reasonable agreement.}   
\label{fig:etg-raz} 
\end{center}
\end{figure}

The distribution of the photon transverse energy for the selected \wg candidate events is shown in Fig.~\ref{fig:etg-raz} (left) with the contribution from the signal and backgrounds shown separately. The $\etg$ distribution with a reference value of anomalous $WW\gamma$ coupling is also shown. The data is found to agree well with the SM signal and background prediction. 
The cross section is estimated using the formula 
\begin{center}
\begin{math}
\sigma\times {\cal BR}(W\gamma\to\ell\nu_{\ell}\gamma) = \frac{N_{events} - N_{bkg}}{{\cal A}\cdot\epsilon\cdot\cal{L}} 
\end{math}
\end{center}
where $N_{events}$ and $N_{bkg}$ are the number of selected \wg candidate events and the number of estimated background events respectively. ${\cal A}$ and $\epsilon$ are the fiducial acceptance of the detector and the efficiencies of the various event selection criteria while ${\cal L}$ is the integrated luminosity of the dataset used in the measurement. 

The cross section for the combined electron and the muon channels is estimated to be 56.3$\pm$5.0~(stat.)~$\pm$5.0~(syst.)$\pm$2.3(lumi.)~pb which is in good agreement with the SM predicted value of 49.4$\pm$3.8~pb. The ratio method gives a systematic uncertainty of 6.3\% and 6.4\% for the electron and muon channels respectively. The photon energy scale uncertainty is 4.2\% for the electron channel and 4.5\% for the muon channel. The uncertainty on the integrated luminosity is 4\%~\cite{Chatrchyan:2011rr}.  

\section{The radiation amplitude zero}
\label{sec:raz}

The radiation amplitude zero or {\it RAZ} is a unique feature of the \wg production in the SM where the amplitude for the production of \wg events vanish for certain angles that the $W$-boson makes with the incoming quark. A convenient variable at hadron colliders for studying the RAZ is $Q_{\ell}\cdot(\eta_{\gamma}-\eta_{\ell})$~\cite{Baur:1994sa}, where $Q_{\ell}$ is the charge of the lepton. This variable shows a dip at zero indicating the presence of the SM RAZ in the \wg production. ATGC destroys the RAZ feature since is depends critically on the SM gauge nature of the $WW\gamma$ coupling. Further, next-to-leading order effects accompanying the \wg production obscures the dip and makes the detection of the RAZ challenging at the LHC. The plot of $Q_{\ell}\cdot(\eta_{\gamma}-\eta_{\ell})$ from data and the SM \wg signal is shown in Fig.~\ref{fig:etg-raz} (right) and show a reasonable agreement, within error estimates. 

\section{Summary}
\label{sec:summary}

This paper presents the first study of the \wg event production at the LHC at centre of mass energy of 7~TeV, made using the CMS detector. A measurement of the \wg cross section is done using the electron and muon decay channels of the $W$-boson and the measured value is found to be in good agreement with the prediction of the standard model. An attempt has also been made to study the radiation amplitude zero feature of the SM \wg production and with the limited data from the first year of the running of the LHC, the data is found to be consistent with the SM, though with a large uncertainty. The measurement of the cross section was one of the most important goals of this analysis and is the first step towards the determination of the $WW\gamma$ couplings.

\end{document}